# Contractile and mechanical properties in epithelia with perturbed actomyosin dynamics

Sabine C. Fischer[1$], Guy B. Blanchard[2], Julia Duque[4], Richard J. Adams[2], Alfonso Martinez Arias[1], Simon D. Guest[3] and Nicole Gorfinkiel[4,*].

[1]Buchmann Institute for Molecular Life Sciences, Department of Biological Sciences, Goethe University Frankfurt, Frankfurt am Main, Germany. [2]Department of Physiology, Development and Neuroscience, University of Cambridge, Cambridge, UK. [3]Department of Engineering, University of Cambridge, Cambridge, UK. [4]Centro de Biología Molecular "Severo Ochoa", CSIC-UAM, Cantoblanco, Madrid, Spain.
*Author for correspondence (email: ngorfinkiel@cbm.csic.es)

## Abstract

Mechanics has an important role during morphogenesis, both in the generation of forces driving cell shape changes and in determining the effective material properties of cells and tissues. Drosophila dorsal closure has emerged as a reference model system for investigating the interplay between tissue mechanics and cellular activity. During dorsal closure, the amnioserosa generates one of the major forces that drive closure through the apical contraction of its constituent cells. We combined quantitation of live data, genetic and mechanical perturbation and cell biology, to investigate how mechanical properties and contraction rate emerge from cytoskeletal activity. We found that a decrease in Myosin phosphorylation induces a fluidization of amnioserosa cells which become more compliant. Conversely, an increase in Myosin phosphorylation and an increase in actin linear polymerization induce a solidification of cells. Contrary to expectation, these two perturbations have an opposite effect on the strain rate of cells during DC. While an increase in actin polymerization increases the contraction rate of amnioserosa cells, an increase in Myosin phosphorylation gives rise to cells that contract very slowly. The quantification of how the perturbation induced by laser ablation decays throughout the tissue revealed that the tissue in these two mutant backgrounds reacts very differently. We suggest that the differences in the strain rate of cells in situations where Myosin activity or actin polymerization is increased arise from changes in how the contractile forces are transmitted and coordinated across the tissue through ECadherin-mediated adhesion. Altogether, our results show that there is an optimal level of Myosin activity to generate efficient contraction and suggest that the architecture of the actin cytoskeleton and the dynamics of adhesion complexes are important parameters for the emergence of coordinated activity throughout the tissue.

Introduction

Mechanical input is increasingly seen as an important regulator of development in both normal conditions and disease. Mechanical forces appear to regulate a variety of basic cellular processes such as cell adhesion, cell signalling, proliferation and differentiation (reviewed in [1,2]). During morphogenesis, mechanics has a role both in the generation of forces important for cell shape changes and cellular rearrangements and in determining the effective mechanical properties of cells and tissues. Cell mechanical properties such as stiffness and viscosity are key cellular parameters that will influence how cells respond to forces and how they will transmit these forces across the tissue. Hence, mechanical properties are crucial in determining the rate at which developmental processes occur and the timing of patterning events [3]. Changes in mechanical properties in embryonic tissues can therefore produce variation at the whole organismal level and contribute to the generation of variability in natural populations.

The mechanical properties of cells can be measured using a variety of experimental techniques that probe the response of different parts of the cells to an externally applied force (reviewed in [4,5]). These rheological studies have shown that a cell's material properties depend on the nature of the applied stress and the timescale over which the response is measured. These studies together with the studies performed in *in vitro* reconstituted networks are starting to reveal that cell mechanical properties largely depend on the architecture and the dynamics of the actomyosin cytoskeleton. Because the same core machinery is involved both in the generation of contractile forces and in determining the stiffness and viscosity of cells and tissues, it is likely that there is a control over these properties to ensure and stabilize the deformations required for proper development.

Despite the vast amount of rheological work performed at the level of single cultured cells, the ability to measure mechanical properties in living embryos still poses an experimental and a theoretical challenge [4,6]. Some exceptions include the measurement of the stiffness of *Xenopus* embryonic tissues by subjecting tissue explants to compression and the measurement of cortical tension in progenitor cells from gastrulating zebrafish embryos [7,8]. These experiments have shown that mechanical stiffness varies among tissues and according to developmental stage. They have also shown that tissue stiffness depends on the activity of the actomyosin cytoskeleton *in vivo*, with decreasing Myosin activity or actin polymerisation decreasing stiffness and cortical tension [7,8,9]. However, how variations in actin and myosin activity affect morphogenesis by dually impinging onto force generation mechanisms and mechanical resistance in living embryos is not known. Experimental manipulation of actin and myosin activity in embryos can help to understand how the balance between movement and resistance is maintained to give rise to wild type morphogenesis. Moreover, it could inform the generation of artificial self-deforming tissues in tissue engineering with specific requirements for stiffness and deformation rates [10].

To gain some insight into how cell and tissue mechanical properties emerge from the activity of the cytoskeleton and cell-cell interactions in tissues undergoing morphogenesis, we made use of the process of dorsal closure (DC) of the *Drosophila* embryo. Half way through embryo development, after the full retraction of the germ-band, the *Drosophila* embryo has a dorsal gap in the epidermis that is bridged by an extra-embryonic epithelium, the amnioserosa (AS) [11]. During DC the gap is closed by coordinated contraction of the AS and extension of the epidermis until the AS disappears inside the embryo and the epidermis is sealed. A supra-cellular actin cable assembled at the interface of the two tissues further contributes to closure [12,13]. Where the two epidermal fronts are sufficiently close, they are engaged in a zippering mechanism which starts at the two canthi of the dorsal gap and progresses inwards, ensuring the correct fusion of the epidermis to a continuous sheet [14,15].

DC provides the opportunity to address general questions about the mechanics of cells and tissues in a manageable system. During the last few years it has become clear that the AS is amenable to genetic and to mechanical perturbations and it is possible to track the behaviour of its approximately 150 cells over time in wild type and mutant embryos as well as after laser ablation. This quantitative data combined with mathematical modelling is providing an integrative understanding of the biomechanical basis of this morphogenetic process (reviewed in [16]). Two time scales define the whole process: AS cell shape fluctuations have a time scale of a few minutes, whereas the time scale of the whole tissue contraction can be detected at time scales of tens of minutes [17,18,19]. Interestingly, recent studies on the mechanics of the AS indicate that its cells combine passive viscoelastic behaviour with active contraction [20,21,22,23], and that both are essential to capture the dynamics of closure [20,21].

In this work we measured cell mechanical properties using laser ablation experiments and explored how these properties impinge onto the rate of deformation of AS cells during DC. We analysed how these properties are affected in embryos in which actomyosin dynamics was perturbed in various ways. We found that a decrease in Myosin phosphorylation induces a more fluid-like behaviour compared to wild type cells. Conversely, an increase in Myosin phosphorylation and an increase in actin polymerization both induce a more solid-like behaviour. Contrary to expectation, these two perturbations of Myosin and actin dynamics have an opposite effect on the strain rate of cells during DC. While an increase in actin polymerization increases the strain rate of AS cells at the onset of DC, an increase in Myosin phosphorylation produces cells that contract very slowly. Quantifying the spatial decay of the influence of laser perturbation revealed that the AS tissue in these two mutant backgrounds reacts differently to the perturbation generated by the laser cut. Live imaging of actomyosin reporters and FRAP experiments in these two mutant situations revealed differences in the organization of the cytoskeleton and in the dynamics of ECadherin. We suggest that the differences in the strain rates of cells in situations where Myosin activity or actin polymerization is increased arise from differences in how the contractile forces are transmitted and coordinated across the tissue through ECadherin-mediated adhesion. Altogether, our results show that there is an optimal level of Myosin activity to generate efficient contraction and suggest that the architecture of the actin cytoskeleton and the dynamics of cell-cell adhesion are important parameters for the emergence of coordinated activity throughout the tissue.

Materials and Methods

*Drosophila* strains and microscopy

The following stocks were used: ubiECad-GFP –hereinafter ECadGFP [24], sGMCA [12,25]; zipperCPTI002907 (available from Kyoto Stock Center); c381GAL4 (an AS driver), UASctMLCK, a constitutive active form of Myosin Light Chain Kinase [26], UASDia$^{CA}$, a constitutive active form of the formin Diaphanous [27] and UASMbsN300, a constitutive active form of the Myosin Binding Subunit of Myosin Light Chain Phosphatase [28]. Stage 12-13 *Drosophila* embryos were dechorionated, mounted in coverslips with the dorsal side glued to the glass and covered with Voltalef oil 10S (Attachem). The AS was imaged at 25-28ºC. using an inverted LSM 510 Meta laser scanning microscope with a 40X oil immersion Plan-Fluor (NA=1.3) objective. The whole AS with ECad-GFP was imaged with an argon laser. 15-16 z sections 1.5µm apart were collected every 30 seconds.

Ablations were performed with an ultrafast Mai Tai DeepSee laser (Spectra-Physics, 2.5W, 80Mhz, pulse < 100 femtosec, at 850 nm) attached to an LSM710 laser scanning microscope with a 63X oil immersion Plan-Apochromat (NA=1.4) objective. To perform ablations, the laser at 50-100% power was targeted to a region of interest of approximately 20µm x 0.2µm x 1µm during 30-50 milliseconds. Images were collected every 50 milliseconds.

FRAP experiments were perfomed using an LSM710 laser scanning microscope with a 63X oil immersion Plan-Apochromat (NA=1.4) objective. A circular region of interest (ROI) (r=0.52µm) was bleached with a 488nm laser beam at 100% power. Images were taken before and after bleaching every 2s for 3 minutes. A 3.2 x 3.2µm reference region was also imaged to take into account photobleaching effects.

Image analysis

The quantification of AS apical cell area was done using automated tracking of the AS cell shapes with custom software written in Interactive Data Language (IDL, Exelis) as described previously [18,29]. The area of the whole AS and the recoil of cell vertices after ablation was tracked automatically, applying Mathematica (Wolfram Research) built-in functions. The number of cells in the AS was counted manually.

Linear regression analysis

All linear regressions were performed with the Mathematica built-in function LinearModelFit (Wolfram Research) which provides the best fit parameters along with test statistics including residuals, confidence intervals and p values for parameter t-statistics. If necessary, the confidence intervals were adjusted with the Bonferroni correction.

Analysis of strain rate and recoil after laser ablation

The cell strain rate (or proportional rate of contraction) was measured as previously described [18,29] and is given by:

$$\frac{A(t+dt) - A(t-dt)}{2A(t)dt}, (1)$$

where $A(t)$ is the apical cell area at time $t$ and $dt$ the time between imaging frames. Average strain rates were calculated for different genotypes, pooling all cells from all embryos of each genotype. Epochs during which there were significant differences in the average rate of area change between any pair of genotypes were calculated as follows. First, a Butterworth filter was applied and oscillatory behaviour with a period of less than 10 minutes was removed. The remaining trends were resampled every two minutes and at each time point we applied a linear mixed model, comparing the strain rate trend data with embryo as a random effect.

To analyse the local response to ablation we used two different frameworks that have been previously used in *Drosophila* tissues. The first one makes use of the observation that over two orders of magnitude of time (0.1 to 10 s), the recoil behaviour is well fit by a weak power law [22]. We grouped the vertices of the first row next to the cut by embryo and fitted a power law $d \cdot t^\alpha$, where $d$ relates to the extent of displacement and $\alpha$ is an indicator of the viscoelastic properties of the cells, which varies between 0 and 1. For materials that follow power-law rheology, a smaller exponent implies more solid-like behaviour while a larger exponent implies more fluid-like behaviour [22].

The second framework is the one that is most widely used and assumes linear viscoelastic behaviour, where the vertices of the cells are connected by Kelvin-Voigt modules, which consist of a spring and a dashpot in parallel, representing the elastic and viscous properties respectively. The dynamics after laser ablation are then described by

$$\eta \dot{x} + \xi x = T, x(0) = 0, \quad (2)$$

where $T$ denotes the tension released through ablation, $\eta$ is the viscous coefficient of the dashpot and $\xi$ the stiffness of the spring. Equation (2) is solved by

$$x(t) = \frac{T}{\xi}\left(1 - e^{-\frac{\xi t}{\eta}}\right) \text{ and } \dot{x}(t) = \frac{T}{\eta} e^{-\frac{\xi t}{\eta}}. \quad (3)$$

The maximal displacement is given by $\lim_{t \to \infty} x(t) = \frac{T}{\xi}$ and $\frac{\eta}{\xi}$ is the decay time constant of the recoil velocity. We obtained these two parameters by fitting a single exponential to the data. The initial recoil velocity $\dot{x}(0) = \frac{T}{\eta}$ was obtained from the movement of a vertex between the last frame before ablation and the frame at three seconds after ablation.

Nonlinear regressions were performed with the Mathematica built-in function NonlinearModelFit (Wolfram Research). The resulting parameter values are shown as mean +/- standard error of the mean. To assess the significance of differences between genotypes we applied the Wilcoxon-rank-sum test with Holm's correction in R (R Foundation for Statistical Computing). Embryos for which at least two of the parameters were more than two standard deviations away from the mean were classified as outliers and removed from the analysis (two embryos for wild type, none for ASGal4/UAS-MbsN300 and one embryo each for ASGAL4/UASctMLCK and ASGAL4/UASDia$^{CA}$). The kernel density estimates were determined with the Mathematica built-in function SmoothKernelDistribution (Wolfram Research).

The tissue response to ablation was characterised by the recoil of the vertices in the first and second row next to the cut. We present the maximal displacement of 2$^{nd}$ neighbour vertices as well as the parameters for fitting the power law $d \cdot t^\alpha$. Vertices that could not be fitted by α between 0 and 1 were excluded.

FRAP analysis

For FRAP analysis, normalized fluorescence over time for each individual experiment was fitted to a simple exponential function of the form: I(t) = A(1-exp(-bt)) using the MATLAB built-in function nlinfit and nlparci (MathWorks, Natick, MA), where A is the mobile fraction and b is $\frac{\ln 2}{\tau_{1/2}}$, where $\tau_{1/2}$ is the half time of the recovery. Mean parameters were calculated for each genotype. To assess the significance of differences between genotypes we applied a one way ANOVA for A and b and a comparison test (MATLAB function multcompare).

Results

<u>Perturbing actin and myosin dynamics alters the strain rate of AS cells</u>

During DC, AS cells progressively contract their apical surface area giving rise to one of the major forces driving the closure of the dorsal epidermis. The proportional rate of apical contraction, or strain rate, of AS cells is an invariant magnitude of the rate of contraction of these cells. We have previously measured the strain rate of AS cells over the course of DC and shown that it increases over time, the dynamics of which can be separated into two main phases, the slow and the fast phase, according to the pace at which the strain rate increases. This contraction is anisotropic, with cells contracting preferentially in the medio-lateral (ML) direction for most of DC with a small contribution of contraction in the antero-posterior (AP) orientation towards the second half of the process [18], To explore how actomyosin dynamics impinge on cell contractility, we measured the total surface covered by the AS, the cell size and the strain rate of AS cells in embryos in which actomyosin activity was perturbed (Fig. 1 and S1 and S2). Specifically, we decreased and increased Myosin phosphorylation in the AS by ectopically expressing a constitutively active form of the myosin binding subunit (MbsN300) of the Myosin Light Chain Phosphatase and a constitutively active form of the Myosin Light Chain Kinase (ctMLCK), respectively. Myosin phosphorylation is crucial to allow single Myosin hexamers to assemble into bipolar and highly processive minifilaments (reviewed in [30]). Thus, increasing or decreasing Myosin phosphorylation is likely to directly affect Myosin motor activity. We also perturbed actin network architecture by increasing actin linear polymerization through the ectopic expression of a constitutive active form of the *Drosophila* formin Diaphanous (Dia$^{CA}$).

In embryos with decreased Myosin activity, the AS covers a larger surface area than in wild type embryos (Fig. 1 and S1), AS cells are initially twice the size of wild type ones (Fig. S1), and are very elongated in the medio-lateral (ML) axis (Fig. 1B). These observations suggest that these cells cannot contract properly and that they are not able to overcome the resistive force of the epidermis. Despite altered cell shapes, the strain rate of these cells was not significantly different from wild type during most of DC, only having a reduced strain rate during late stages of the process (Fig. 1E). Looking at the contribution of ML and AP strain rate to the total cell area strain rate, we observe that ML contraction is stronger than in wild type embryos during the first 40min of DC (Fig. 1F). However, since this difference is lost when looking at total cell area contraction it is likely that these cells are contracting in ML but also expanding in AP during the first part of DC (although we were not able to detect significant differences in AP strain rate between wild type and ASGal4/UASMbsN300 embryos [Fig. 1G]).

Conversely, the over-activation of Myosin activity in the AS through expression of a constitutive active form of Myosin Light Chain Kinase (MLCK) and increased actin filament elongation through expression of the formin Diaphanous in the AS produces precocious contraction of AS cells (Fig. 1C, D) [17,31]. This is reflected in the reduced surface area covered by the AS and the smaller apical surface area of these cells, which is approximately 50% smaller than in wild type embryos (Fig. S1). Interestingly, the strain rate of AS cells in these two genotypes shows distinguishable dynamics: AS cells in ASGal4/UASctMLCK embryos contract slower than wild type cells from 40 minutes onwards (Fig. 1E). This observation shows that, contrary to intuition, an increase in Myosin phosphorylation results in a decrease in the strain rate of AS cells. In contrast, in ASGal4/UASDia$^{CA}$ embryos cells start contracting faster than wild type cells, and their rate of contraction remains constant during most of the process. Hence, from about 70 minutes onwards they contract more slowly than wild type AS cells (Fig. 1E). This faster strain rate results from a contribution of contraction both in the ML and AP direction (Fig. 1F, G). Note that initially only ASGal4/UASDia$^{CA}$ embryos show a significant AP contraction (see below).

Altogether, the differences in AS surface area, cell size and strain rate among the genotypes suggest that the contractile machinery is disrupted in different ways. Since the same cytoskeletal machinery underlies the contractile activity and the passive mechanical properties of the cells, we sought to investigate how actomyosin perturbation impinges onto the mechanical properties of cells and tissue in these different genotypes.

<u>Mechanical perturbation of the AS</u>

Laser ablation is emerging as a powerful tool to analyse the balance of stresses within cells and their mechanical properties in live embryos [32]. In the AS, tissue wide cuts were performed to infer the tissue level forces involved in DC [12,33] and ablations targeted to single cells shed light onto how the mechanical properties of its constituent cells evolve during the process [22]. In this work we are interested in the average cell behaviour and global mechanical properties of the cells. To avoid the known differences in recoil behaviour after ablation at cell centres and cell edges [22], we conducted ablations of the whole apical cellular cortex spanning 3-4 cells and analysed the recoil response of the surrounding tissue after ablation. Laser cuts of approximately 20 μm were mostly performed oriented parallel to the anterior-posterior axis and thus perpendicular to the axis of major contraction during the slow phase of DC (Fig. 2A-D, Supplementary Movies 1-4).

*Local response to ablation*

We obtained the temporal evolution of local recoil behaviour by tracking the vertices of the first row of cells above and below the laser cut. The displacement over time shows a characteristic signature, such that initially it increases rapidly and then levels off after a few seconds. To infer the mechanical properties of the cells from such measurements we applied two approaches that have previously been used in the literature (see Materials and Methods). Fitting a power law to the recoil displacement data from wild type and genetically perturbed embryos (Fig. 2E,G and Fig. S5) showed differences in the mechanical properties of cells between the different genotypes. For wild type embryos, we obtained a power-law that is in agreement with the one previously reported [22]. We found that there is a gradation in the power-law exponent for the different genotypes analyzed: ASGal4/UASMbsN300 embryos show the highest power-law exponent followed by wild type embryos, and finally ASGal4/UASctMLCK and ASGal4/UASDia$^{CA}$ embryos with the smallest power-law exponent (Fig. 2G). This result shows that cells with increased levels of phosphorylated Myosin or with an increased elongation of actin filaments are more solid-like than wild type cells. In contrast, cells with decreased Myosin phosphorylation levels are more fluid-like.

We also analysed the recoil behaviour of the tissue using a simple viscoelastic system or Kelvin-Voigt model (Fig. 2F and Fig. S3) that can provide insights into the relation between three important mechanical properties; tension, viscosity and stiffness [34,35]. Cell vertices surrounding the site of a laser cut exhibit a recoil displacement characterized by an initial outward velocity that decays exponentially over time. The decay time constant of the recoil velocity is a signature of the mechanical properties of the cells, in particular it is the ratio of the viscosity to the stiffness of the cells. The decay time constant of ASGal4/UASctMLCK and ASGal4/UASDia$^{CA}$ embryos is smaller than the wild type one (Fig. 2H), suggesting a higher stiffness relative to viscosity, in agreement with the power-law exponent. In contrast, ASGal4/UASMbsN300 embryos show a higher decay time constant compared to wild type (Fig. 2H), suggesting a higher viscosity relative to stiffness, also in agreement with the power-law exponent. Interestingly, these results show that the same changes in the mechanical properties upon perturbation of the actomyosin cytoskeleton can be detected independently of the framework chosen to measure the mechanical behaviour.

Another important parameter that can be measured within the Kelvin-Voigt framework is the maximal displacement after ablation, which is proportional to the tension present on the ablated region just before ablation in a manner that depends on the stiffness (maximal displacement = tension/stiffness). We find that the maximal displacement is different among the different genotypes analyzed (Fig. 2I). Wild type embryos show the highest maximal displacement followed by ASGal4/UASMbsN300 and ASGal4/UASctMLCK embryos, while AS cells from ASGal4/UASDia$^{CA}$ embryos have the smallest maximal displacement. Finally, we also measured the initial recoil velocity ($v_0$) by computing the displacement of the vertices during the first seconds after the ablation (Fig. 2J), which in the Kelvin-Voigt framework is proportional to the tension present just before ablation divided by the viscosity. Wild type embryos show the highest initial recoil velocity followed by ASGal4/UAS-ctMLCK embryos, while ASGal4/UAS-MbsN300 ASGal4/UAS-dia$^{CA}$ embryos have the smallest initial recoil velocity.

These measurements do not yield absolute values for the stiffness, viscosity or the tension in the system but rather ratios between any two of these. Therefore, when computing these parameters in embryos where the actomyosin cytoskeleton has been perturbed in various ways, differences in the maximal displacement, the decay kinetics or the initial recoil velocity between different genotypes cannot be attributed to a change in an individual mechanical property. However, they unambiguously show that a decrease in Myosin activity yields cells that are more fluid-like, where viscosity predominates over stiffness when compared to the wild type, while an increase in Myosin activity and in actin polymerization yields cells that are more solid-like, where stiffness predominates over viscosity.

Our results also reveal a complex relationship between cell shape, strain rate and local mechanical properties. For example, a decrease in Myosin phosphorylation produces cells that are more fluid-like and probably more compliant. This could explain the extreme elongated shape of these cells, stretched in ML by the resistive force of the epidermis. However, these cells do not contract slower than wild type cells for most of DC, suggesting that in these cells there is enough Myosin activity to drive contraction or that other compensatory mechanisms are driving AS contraction. On the other hand, embryos with increased Myosin phosphorylation or with increased actin polymerization show a more solid-like behaviour than the wild type, AS cells are much smaller than wild type ones but their rate of contraction deviate from the wild type in different ways (Fig. 1). A possible source of this difference between local mechanical properties and strain rate values could come from the different temporal scales at which these observables are measured. In the laser ablation experiments, the behaviour of the system is assessed at short time scales ranging 0.1-10 seconds and cannot be assessed for longer time scales because cells start to actively remodel their cytoskeleton through wound healing processes. In contrast, the strain rate of AS cells was measured over longer time scales (> 5 minutes) and because it is an average of the rate of contraction of all the cells in the tissue it represents a tissue property more than a local cell property.

*Tissue response to ablation*

Thus, to gain insights into tissue mechanical properties, we explored the overall tissue response to ablation. For this, we measured the maximal displacement of vertices surrounding the cut (1$^{st}$ neighbour vertices) and of 2$^{nd}$ neighbour vertices. We analysed the recoil response of 2$^{nd}$ neighbour vertices as well as how the maximal displacement reduces with distance from the laser cut. Thus, these measurements give us an estimate of how the perturbation generated by the cut is transmitted across the tissue and can give an idea of the overall tissue behaviour.

We fitted a power law to the recoil displacement of 2$^{nd}$ neighbour vertices from wild type and genetically perturbed embryos. We then compared the prefactor d, a parameter that reflects the extent of the recoil, and the power-law exponent, a parameter describing how solid-like the material is, between 1$^{st}$ and 2$^{nd}$ neighbour vertices for the same genotype (Fig. 3). In wild type, ASGal4/UASMbsN300 and ASGal4/UASctMLCK embryos there is a clear decrease in the prefactor d of 1$^{st}$ and 2$^{nd}$ neighbour vertices, which reflects a reduction in the extent of the recoil (Fig. 3A-C, E). However, there is no such decrease in the prefactor d when analyzing the recoil kinetics of 1$^{st}$ and 2$^{nd}$ neighbour vertices of ASGal4/UASDia$^{CA}$ embryos (Fig. 3D,E). This suggests that the perturbation generated by the laser cut is damped across the tissue in wild type, ASGal4/UASMbsN300 and ASGal4/UASctMLCK embryos but it does not do so in ASGal4/UASDia$^{CA}$ embryos. In contrast, the power-law exponent alpha, does not decay from 1$^{st}$ to 2$^{nd}$ neighbour vertices in any of the genotypes analysed, showing that mechanical properties of cells are an invariant parameter that does not change across the tissue (Fig. 3A-D, F).

We also plotted the maximal displacement of 1$^{st}$ and 2$^{nd}$ neighbour vertices as a function of the distance from the wound (Fig. 4). Although it has been suggested that the displacement of vertices decays as 1/r, this represents an idealized situation where the tissue sheet would undergo infinitesimal deformations [22]. Our data is well fitted by a linear regression (Fig. S4) and this provides the "decay spatial constant" (slope of the linear regression), a useful parameter that characterises how far the effect of the mechanical perturbation spreads across the tissue. These results confirm the results from the power law fitting of $2^{nd}$ neighbour vertices showing that the maximal displacement of vertices after ablation decays with increasing distance from the wound in all except ASGal4/UASDia$^{CA}$ embryos (Fig. 4A-E). We asked if this absence of decay in the latter could be due to the small maximal displacement observed in ASGal4/UASDia$^{CA}$ embryos. To test this possibility, we performed laser ablations with increased laser power in both ASGal4/UASctMLCK and ASGal4/UASDia$^{CA}$ embryos (Fig. S5). We observed that although the maximal displacement of $1^{st}$ and $2^{nd}$ neighbor vertices is increased in both genotypes, ASGal4/UASctMLCK embryos still show a damping of the maximal displacement with increasing distance from the wound while in ASGal4/UASDia$^{CA}$ embryos there is no such a decay (Fig. S5). Moreover, we found that such behavior is also observed when performing cuts in the ML axis, showing that mechanical perturbation is transmitted across the tissue in an isotropic way in ASGAL4/UASDia$^{CA}$ embryos.

Altogether, these results suggest that in wild type, ASGal4/UASMbsN300 and ASGal4/UASctMLCK embryos the perturbation generated by the laser cut is damped across the tissue and does not spread further than the first neighbours of the cells affected by the cut. In contrast, in ASGal4/UASDia$^{CA}$ embryos the perturbation generated by the ablation cut is not damped with increasing distance from the wound and spreads isotropically across the tissue. Interestingly, these results correlate with the high cell strain rate in these embryos in both ML and AP directions and thus strongly suggest a correlation between tissue mechanical properties and strain rate. While in ASGal4/UASctMLCK embryos cells would not be able to transmit local tension across neighbours, giving rise to a slow contracting tissue, in ASGal4/UASDia$^{CA}$ embryos cells would be able to transmit locally generated contractile tension over long distances, isotropically, giving rise to a rapidly contracting tissue.

To better understand how these differences in tissue behaviour could arise, we explored the organization of the cytoskeleton and the dynamics of adhesion in these different mutant situations.

*Cytoskeletal and adhesion dynamics in embryos with perturbed actomyosin activity*

We explored the cellular basis of the differential behaviour exhibited by the AS of ASGal4/UASctMLCK and ASGal4/UASDia$^{CA}$ embryos by performing time-lapse movies of wild type and perturbed embryos carrying the actin reporter sGMCA, which is the actin binding domain of the Moesin protein fused to GFP, and thus reveals the dynamic localization of F-actin, and with the non-muscle Myosin II Heavy Chain reporter, zipperYFP. We observed significant differences in the subcellular localization of the actomyosin cytoskeleton and in the overall organization of the tissue in the different genotypes (Fig. 5 and Movies 5-7). AS cells from ASGal4/UASDia$^{CA}$ embryos did not exhibit distinctive cell area fluctuations [17]. However, they exhibit a dynamic medial actin population and the formation of clear transient myosin foci (Movies 6a, 6b). These cells also show increased levels of actin at the level of cell-cell junctions that extends basolaterally, which seem to recruit moderate levels of Myosin (Fig. 5B, E-E"). These observations show the existence in these cells of a very stable junctional/cortical actin network that could contribute to the solid-like properties of the cells and could give rise to a highly coherent tissue, allowing for the rapid transmission of contractile forces.

In contrast, AS cells from ASGal4/UASctMLCK embryos show discernible cell area fluctuations [17]. These cells exhibit dynamic foci of actin in the medial region, which are increased in number and look more dense and compact than the ones observed in wild type embryos (Fig. 5C and Movie

7a). In time-lapses some of these foci assemble and disassemble in situ, probably associated with bleb contraction, while others flow across the apical surface of cells. Actin localization at the level of cell-cell junctions is much lower than in ASGal4/UASDia$^{CA}$ embryos (Fig. 5C). Imaging of zipperYFP showed that Myosin accumulates at the level of cell-cell junctions (Fig. 5F") but since this accumulation is not related to an increase of junctional actin, it is likely that this Myosin population cannot sustain strong tension. The medial Myosin network did not form unique foci in individual cells but instead accumulated in multiple small mobile clusters (Fig. 5F-F" and Movie 7b). These observations suggest to us that although the medial actomyosin network in these cells can generate local contraction associated with cell fluctuations and the presence of blebs, the absence of a junctional actin population prevents the translation of this local contractile activity into a rapid contracting tissue.

To test if the presence of a strong junctional actin population in ASGal4/UASDia$^{CA}$ embryos impacts adhesion dynamics we measured ECadherin turnover by performing fluorescence recovery after photobleaching (FRAP) experiments. We conducted FRAP assays on ECadherinGFP in wild type (Fig. 6A, B), ASGal4/UASMbsN300, ASGal4/UASctMLCK and ASGal4/UASDia$^{CA}$ (Fig. 6B) embryos during the slow phase of DC. Interestingly, only ASGal4/UASDia$^{CA}$ embryos show a significantly different mobile fraction and half-time recovery of ECadherin. In particular, these embryos show a drastic increase in the immobile fraction, showing that junctional ECadherin is more stable upon activation of Dia. Interestingly, ECadherin stability has been associated with increased adhesion (reviewed in [36]). These results thus show that junctional actin stabilizes ECadherin and suggest that adhesion is an important parameter for the transmission of locally-generated contractile forces to the scale of the whole tissue.

Discussion

In this work, we have analysed the contractile and mechanical properties of epithelial cells undergoing morphogenesis using a multi-disciplinary approach that combines kinematic descriptions of cell shape changes, genetic perturbation, laser ablation and cell biology. Our results reveal a complex relationship between cell shape, contractile activity and mechanical properties in living tissues. Previous rheological studies both in cell culture and in tissue explants have shown that the state of the actomyosin cytoskeleton is a key factor determining the contractile and mechanical properties of cells. Our laser ablation experiments and analysis also show that mechanical properties of cells change upon perturbation of actomyosin dynamics in living embryos. We found that a decrease in Myosin phosphorylation makes the cells more fluid-like, suggesting an increase in viscosity and/or a decrease in stiffness. Although these cells do not show a significantly slower rate of contraction, they are much larger than wild type cells and their shape is likely to result from extrinsic forces acting on a more compliant material. In contrast, both an increase in Myosin phosphorylation and an increase in actin polymerization make the cells more solid-like, showing an increase in stiffness relative to viscosity compared to the wild type tissue. These observations are in agreement and also extend previous results obtained from *in vitro* reconstituted networks, cultured cells and tissue explants [8,37,38]. For example, actin networks connected by flexible cross-linkers have been shown to stiffen with the length of the filaments [39]. The longer the actin filament, the more crosslinkers are bound and simultaneously prevent the actin filament from deforming. Similarly, Myosin motors pulling on a cross-linked actin network also increase the stiffness of the network [38].

The similarities in the cell response to laser ablation between ASGal4/UASctMLCK and ASGal4/UASDia$^{CA}$ are lost when comparing the cell strain rate during early stages of DC, which is particularly high for the latter. This faster rate of contraction results from a contraction in both ML and AP axes. Interestingly, we observed that the tissue response to the perturbation generated by the laser cut is clearly different between the two genotypes. While in ASGal4/UASctMLCK embryos the perturbation generated by laser ablation decays rapidly with increasing distance from the wound, in ASGal4/UASDia$^{CA}$ embryos the perturbation only very slowly fades with distance and this effect is evident in both AP and ML orientations. We suggest that this differential tissue behaviour arises from differences in the organization of the cytoskeleton and the stability of cell-cell junctions in these two genetic backgrounds. In ASGal4/UASDia$^{CA}$ embryos, a strong mechanical coupling between AS cells, due to the presence of a junctional actin population and reinforced intercellular junctions, gives rise to a rapidly contracting tissue. A role for actin architecture in stabilizing and strengthening adhesion has been observed in several systems (reviewed in [36]) but it is also worth noting that tension per se has been shown to strengthen ECadherin-mediated adhesion [40]. In ASGal4/UASctMLCK embryos, although high Myosin levels suggest that local tension is high, ECadherin is not stabilized and contractile forces cannot be efficiently transmitted across the tissue. Interestingly, recent work on *in vitro* reconstituted networks shows that increasing Myosin levels beyond an upper threshold produces ruptures across the actin network [41], suggesting that to coordinate contractions over mesoscopic length scales, motor activity should not increase excessively. Our work also shows that adhesion plays an essential contribution when considering the mechancial properties of epithelia.

It is worthwhile considering where wild type cells and tissue sit in the landscape of possible combinations of cell mechanical properties and strain rate. Considering the relaxation time constant obtained from laser ablation experiments (which measures whether the material is more solid- or fluid-like) and the strain rate from kinematic measures, we can plot approximate locations of the genotypes on two independent axes for these parameters (Fig. 7). This suggests that the wild type is poised in the middle of the genotypes. Based on our work and other published data [22], we suggest that as the embryo progresses through the different phases of DC, there is a transition towards a more solid-like tissue as the contractile strain rate increases. We propose that this transition cannot be achieved through an increase of Myosin activity only but has to be accompanied by a change in the architecture and dynamics of the actin cortex that facilitates the mechanical coupling of cells. In agreement with this idea, it has recently been shown that during mesoderm invagination, Dia-mediated actin polymerization maintains ECadherin at the level of cell junctions, allowing contractile forces to be transmitted across the tissue [42].

Acknowledgements

We are very grateful to Alexandre Kabla, Pedro F. Machado and Joaquín de Navascués for fruitful discussions. We also thank the Confocal Facility at the Centro de Biología Molecular (SMOC) for their help in setting up the conditions for laser ablation. We thank the following funding bodies for their support: Engineering and Physical Sciences Research Council (SF), Biotechnology and Biological Sciences Research Council grant (No. BB/J010278/1) to RJA and Bénédicte Sanson (GBB), Physics of Medicine Initiative (SF), Ministerio de Ciencia e InnovaciónNG, BFU2011-25828 and "Ramón y Cajal" fellowship award) and a Marie Curie Career Integration Grant (NG, PCIG09-GA-2011-293479).

Figure Legends

Figure 1: Cell and tissue kinetics of AS in wild type embryos and embryos with perturbed actomyosin dynamics. (A-D) Still images from time-lapse movies of example embryos of wild type -ASGal4, ECadGFP- (A), ASGal4/UASMbsN300 (B), ASGal4/UASctMLCK (C), and ASGal4/UASDia$^{CA}$ (D) at the onset of AS contraction and after 50 and 100 minutes of contraction. Anterior is to the left in these images. Embryos were staged by comparing the morphogenesis of the posterior spiracles. Average cell area strain rate (E), ML strain rate (F) and AP strain rate (G) over time for wild type (black, pooled from five embryos), ASGal4/UASMbsN300 (green, pooled from five embryos), ASGal4/UASctMLCK (red, pooled from four embryos), and ASGal4/UASDia$^{CA}$ embryos (blue, pooled from five embryos). The widths of ribbons straddling average strain rates represent a measure of combined within- and between-embryo variance. Parts of ribbons drawn in darker shades represent epochs where genotype behaviour was significantly different from wild type. Pairwise comparisons between genotype are shown in Fig. S2.

**Figure 2: Cell mechanical properties of wild type embryos and embryos with perturbed actomyosin dynamics.** (A-D) Still images from time-lapse movies of example embryos of wild type –ECadGFP- (A), ASGal4/UASMbsN300 (B), ASGal4/UASctMLCK (C), and ASGal4/UASDia$^{CA}$ (D) before and 15 s after laser ablation as an overlay comparing the cell boundaries before (green) and after ablation (purple). Cut size was approximately 20 μm, spanning 3-4 cells, as indicated by the scale bar in the pre-ablation images. (E-F) Recoil behaviour of vertices of cells next to the wound for one example embryo fitted by a power law (E) and an exponential (F) for embryos of wild type (black, four vertices), ASGal4/UASMbsN300 (green, six vertices), ASGal4/UASctMLCK (red, two vertices), and ASGal4/UASDia$^{CA}$ (blues, nine vertices). The experimental data is shown as mean values with error bars representing the standard error of the mean. The lines show the fitted curves. (G-J) Parameter values describing the local response to laser ablation. We present the mean value and the standard error of the mean for the power law exponent α (G), the maximal displacement (H), the time decay constant (I), and the initial recoil velocity (J). Stars indicate significant differences ( * p<0.05, ** p<0.01). The data was pooled from 30 embryos (143 vertices) for wild type, 15 embryos (68 vertices) for ASGal4/UASMbsN300, 17 embryos (77 vertices) for ASGal4/UASctMLCK and 24 embryos (144 vertices) for ASGal4/UASDia$^{CA}$. The details of the data analysis are given in Materials and Methods. The distributions of the parameters are shown in Fig. S3.

Figure 3: **Tissue scale recoil kinetics of wild type embryos and embryos with perturbed actomyosin dynamics.** (A-D) Power law prefactor and exponent obtained from fitting the recoil of 1$^{st}$ and 2$^{nd}$ neighbour cell vertices surrounding the wound as a function of their distance to the wound for wild type –ECadGFP- (30 embryos/165 vertices) (A), ASGal4/UASMbsN300 (24 embryos/92 vertices) (B), ASGal4/UASctMLCK (17 embryos/87 vertices) (C), and ASGal4/UASDia$^{CA}$ embryos (24 embryos/128 vertices) (D). The parameters of the vertices surrounding the laser cut and 2$^{nd}$ neighbours are shown as disks and triangles, respectively. The decay spatial constant for power law prefactor d (E) and power law exponent α (F) are shown as mean with error bars indicating a 95% confidence interval. Stars indicate significant difference from 0 with p<0.05. The power law was fitted to each vertex individually. Details of the data fitting and the statistical testing are given in Materials and Methods.

Figure 4: **Tissue mechanical properties of wild type embryos and embryos with perturbed actomyosin dynamics.** (A-D) Maximal displacement after laser ablation of cell vertices surrounding the wound as a function of their distance to the wound for wild type –ECadGFP- (30 embryos/175 vertices) (A), ASGal4/UAS-MbsN300 (24 embryos/115 vertices) (B), ASGal4/UAS-ctMLCK (17 embryos/101 vertices) (C), and ASGal4/UAS-Dia$^{CA}$ embryos (24 embryos/167 vertices) (D). Note the difference in x-axis scale between A, B and C, D. The maximal displacement of the vertices surrounding the laser cut and 2$^{nd}$ neighbours are shown as disks and triangles, respectively. The decay spatial constant is shown as mean with error bars indicating a 95% confidence interval (E). Stars indicate significant difference from 0 (*p<0.05, **p<0.01). Details of the data fitting and the statistical testing are given in Materials and Methods.

Figure 5: Actomyosin localization in **wild type embryos and embryos with perturbed actomyosin dynamics.** (A-C) Still images from a time-lapse movie of wild type –ECadGFP- (A), ASGal4/UASDia$^{CA}$ (B), and ASGal4/UASctMLCK (C) embryos carrying the sGMCA reporter. Note the brighter and denser actin foci in ASGal4/UASctMLCK (arrows) and the low levels of junctional actin. (D-F") Still images from a time-lapse movie of wild type (D-D"), ASGal4/UASctMLCK (E-E"), and ASGal4/UASDia$^{CA}$ (F-F") embryos carrying the zipperYFP reporter. Note the fragmented appearance of Myosin foci in ASGal4/UASctMLCK embryos (arrows). (D'-F') ECadGFP channel. (D"-F") zipperYFP channel.

**Figure 6: ECadherin dynamics in wild type embryos and embryos with perturbed actomyosin dynamics..** (A) Still images from a FRAP experiment of an example cell from a ECadGFP embryo. (B) FRAP curves fitted by an exponential for one example cell of a wild type –ECadGFP- (black), ASGal4/UASMbsN300 (green), ASGal4/UASctMLCK (red), and ASGal4/UASDia$^{CA}$ embryo (blue). The mean +/- standard error of the mobile fraction (C) and the half time recovery (D) were obtained from fitting exponentials to data from 9 cells of 5 wild type embryos, 8 cells of 3 ASGal4/UASMbsN300 embryos, 8 cells of 4 ASGal4/UASctMLCK embryos and 7 cells of 4 ASGal4/UASDia$^{CA}$ embryos. The stars indicate significant differences (*$p<0.05$).

Figure 7: Phase diagram of strain rate and decay time constant. Cell contraction rate (y axis) and cell mechanical properties (x axis) of the different wild type and mutant situations. The results presented in this work show that ASGal4/UASDia$^{CA}$ embryos show fast AS cell contraction rate and cells are more solid-like than wild type cells, ASGal4/UASctMLCK embryos show slow cell contraction rate and cells are also more solid-like than wild type cells, and ASGal4/UASMbsN300 embryos show low cell contraction rate and cells are more fluid-like than wild type cells. Based on the results from this work and from previous published work (Ma et al., 2009), we propose that wild type embryos increase their cell contraction rate and become more solid-like during DC.

Supplementary Information

**Figure S1: Cell and tissue behaviour of the whole AS**. (A) Apical cell area over time for wild type –ASGal4, ECadGFP- (black, pooled from five embryos), ASGal4/UASMbsN300 (green, pooled from five embryos), ASGal4/UASctMLCK (red, pooled from four embryos), and ASGal4/UAS-Dia$^{CA}$ embryos (blue, pooled from five embryos). Shaded regions represent the standard error of the mean. (B) Mean absolute area of the AS over 100 min. Shaded regions represent the standard error of the mean. (data for wild type was pooled from five embryos, for ASGal4/UASMbsN300 from five embryos, for ASGal4/UASctMLCK from four embryos, for ASGal4/UASDia$^{CA}$ from three embryos) (C) Overlay of dotted lines representing the mean area normalised to the mean area at time 0 min for 60 min (slow phase) and fitted lines with mean prediction bands of confidence level 0.95. (D) Slopes of the fitted lines in B presented as mean with 99 % confidence intervals adjusted with the Bonferroni method. Significant differences were determined by comparison of the confidence intervals and are indicated by stars (**$p<0.01$). (E) Number of cells in the AS. Each triangle represents one embryo. Significant differences were determined by a Student's t-test with a Holm's correction and are indicated by stars (**$p<0.01$).

**Figure S2: Pairwise comparisons** of average cell strain rates between the different genotypes. Pairwise comparisons of total strain rate (A-C), ML strain rate (D-F), and AP strain rate (G-I) between ASGal4/UASMbsN300 and ASGal4/UASctMLCK embryos (A, D, G), ASGal4/UASMbsN300 and ASGal4/UASDia$^{CA}$ embryos (B, E, H) and between ASGal4/UASctMLCK, and ASGal4/UASDia$^{CA}$ embryos (C, F. I) (data for wild type was pooled from five embryos, for ASGal4/UASMbsN300 from five embryos, for ASGal4/UASctMLCK from four embryos, for ASGal4/UASDia$^{CA}$ from five embryos). Shaded ribbons straddling average strain rates represent a measure of combined within- and between-embryo variance, while darker shaded ribbons represent epochs where genotype behaviour was significantly different from wild type.

**Figure S3: Parameters** for local response to laser ablation. We show the kernel density estimates for the power law exponent (A), the maximal displacement (B), the time decay constant (C), and the initial recoil velocity (D).

**Figure S4: Residuals** for the fitting of the tissue scale behaviour of wild type embryos and embryos with perturbed actin or myosin dynamics. The random distributions of the residuals (difference between the fitted curve and the data) indicate that the linear curve fits the different parameters and genotypes equally well.

**Figure S5: Tissue** scale behaviour. Maximal displacement after laser ablation performed with increased laser power of cell vertices surrounding the wound as a function of their distance to the wound for cuts parallel to the anterior-posterior axis in ASGal4/UASDia$^{CA}$ embryos (17 embryos/99 vertices) (A), cuts parallel to the dorsal-ventral axis in ASGal4/UASDia$^{CA}$ (12 embryos/71 vertices) (B), and cuts parallel to the anterior-posterior axis in ASGal4/UASctMLCK (21 embryos/ 100 vertices) (C). Note the difference in y-axis range between A, B and C. The maximal displacement of the vertices surrounding the laser cut and $2^{nd}$ neighbours are shown as disks and triangles, respectively. (D) The decay spatial constant is shown as mean with error bars indicating a 95% confidence interval. Stars indicate significant difference from 0 (**$p<0.01$).

Movie 1. Time-lapse movie of laser ablation during dorsal closure in a wild-type embryo carrying the ubiECadGFP transgene (Fig. 4A). The time interval between frames is 0.4 seconds.

Movie 2. Time-lapse movie of laser ablation during dorsal closure in a ASGal4/UASMbsN300 embryo carrying the ubiECadGFP transgene (Fig. 4B). The time interval between frames is 0.5 seconds.

Movie 3. Time-lapse movie of laser ablation during dorsal closure in a ASGal4/UASctMLCK embryo carrying the ubiECadGFP transgene (Fig. 4C). The time interval between frames is 0.3 seconds.

Movie 4. Time-lapse movie of laser ablation during dorsal closure in a ASGal4/UASDia$^{CA}$ embryo carrying the ubiECadGFP transgene (Fig. 4D). The time interval between frames is 0.3 seconds.

Movie 5a. Time-lapse movie of a wild type embryo carrying the SGMCA reporter. The time interval between frames is 10 seconds.

Movie 5b. Time-lapse movie of a wild type embryo carrying the ubiECadGFP and zipperYFP reporters. The time interval between frames is 15 seconds.

Movie 6a. Time-lapse movie of a ASGal4/UASDia$^{CA}$ embryo carrying the SGMCA reporter. The time interval between frames is 10 seconds.

Movie 6b. Time-lapse movie of a ASGal4/UASDia$^{CA}$ embryo carrying the ubiECadGFP and zipperYFP reporters. The time interval between frames is 15 seconds.

Movie 7a. Time-lapse movie of a ASGal4/UASctMLCK embryo carrying the SGMCA reporter. The time interval between frames is 10 seconds.

Movie 7b. Time-lapse movie of a ASGal4/UASctMLCK embryo carrying the ubiECadGFP and zipperYFP reporters. The time interval between frames is 15 seconds.

**Figure 1**

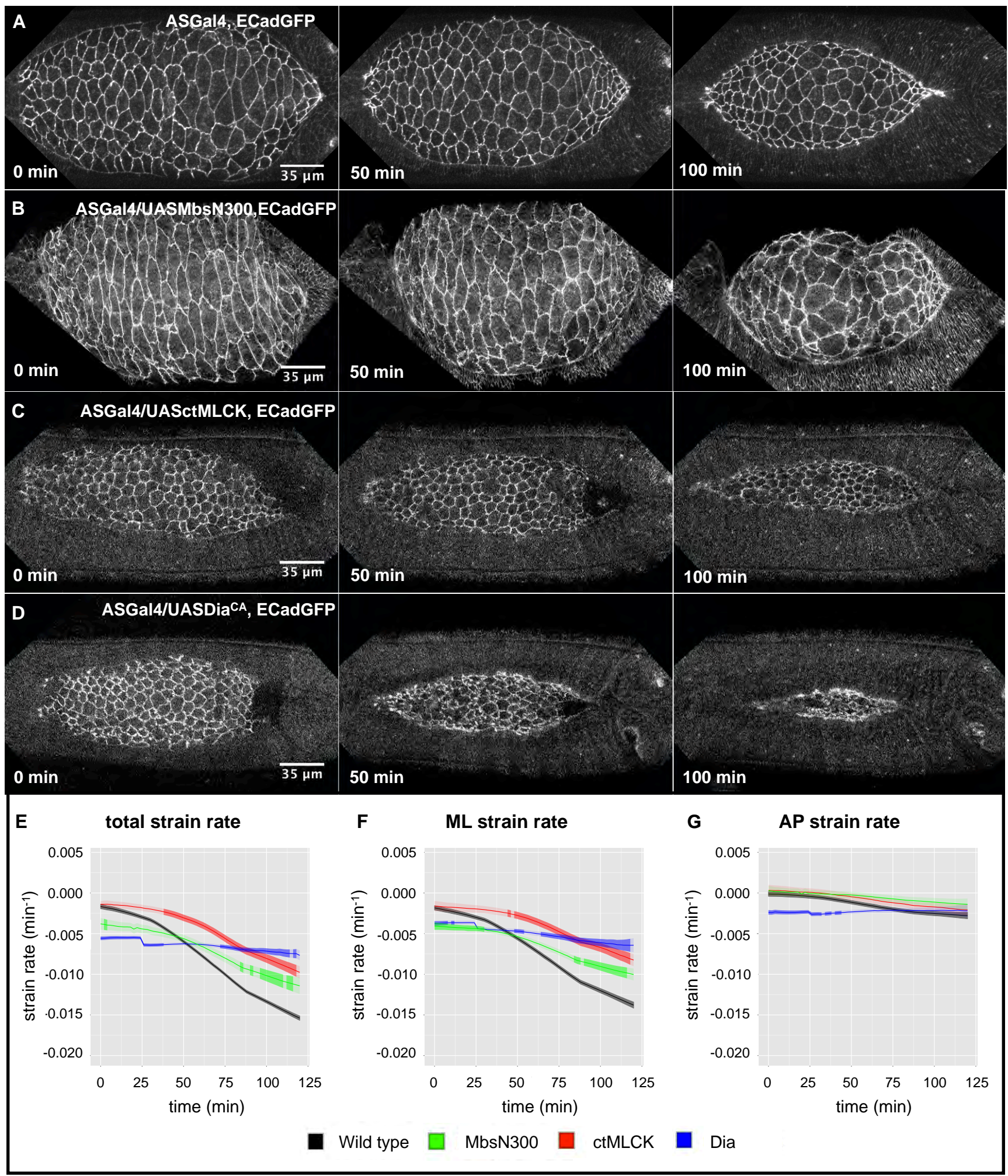

Figure 2

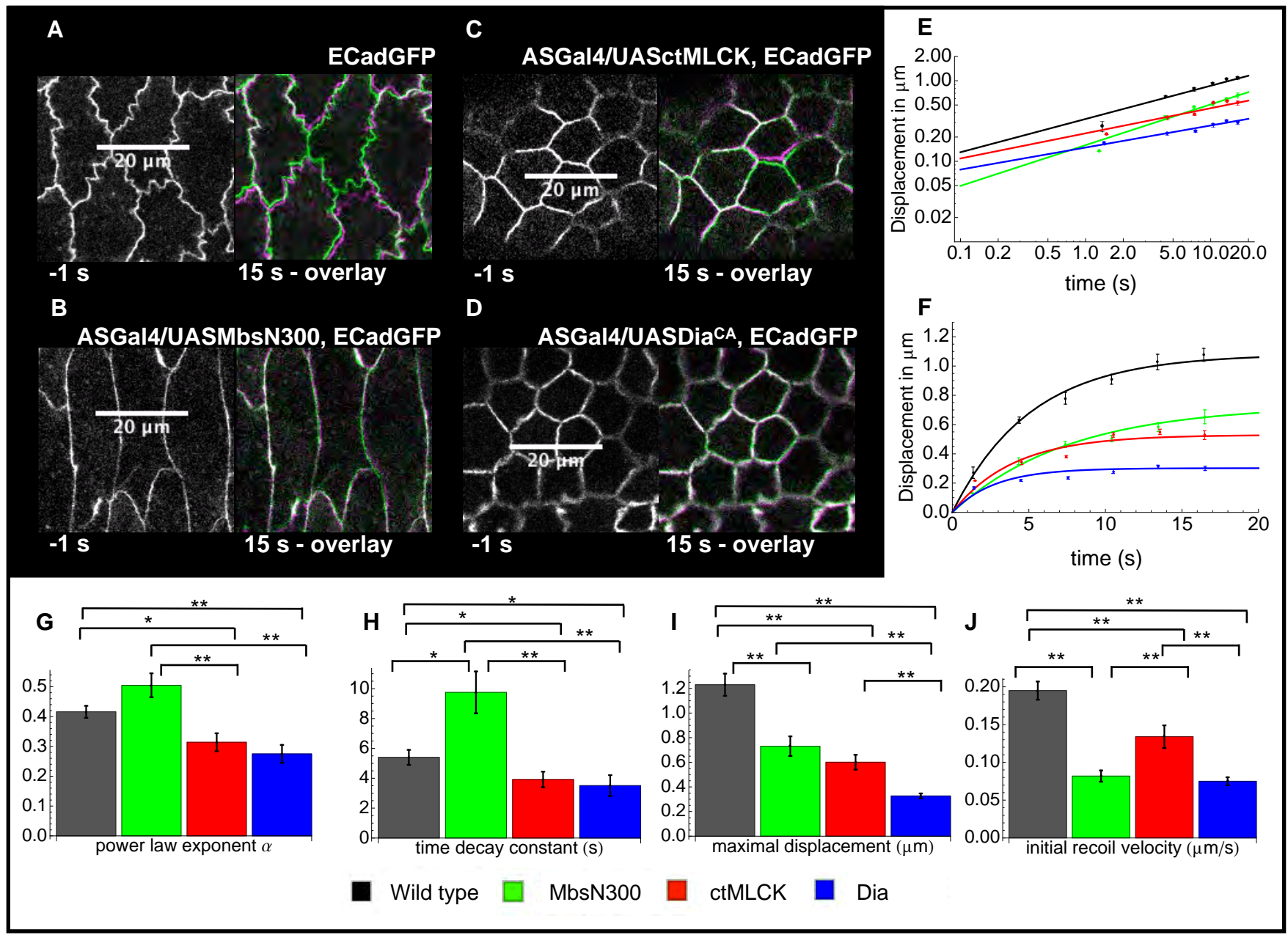

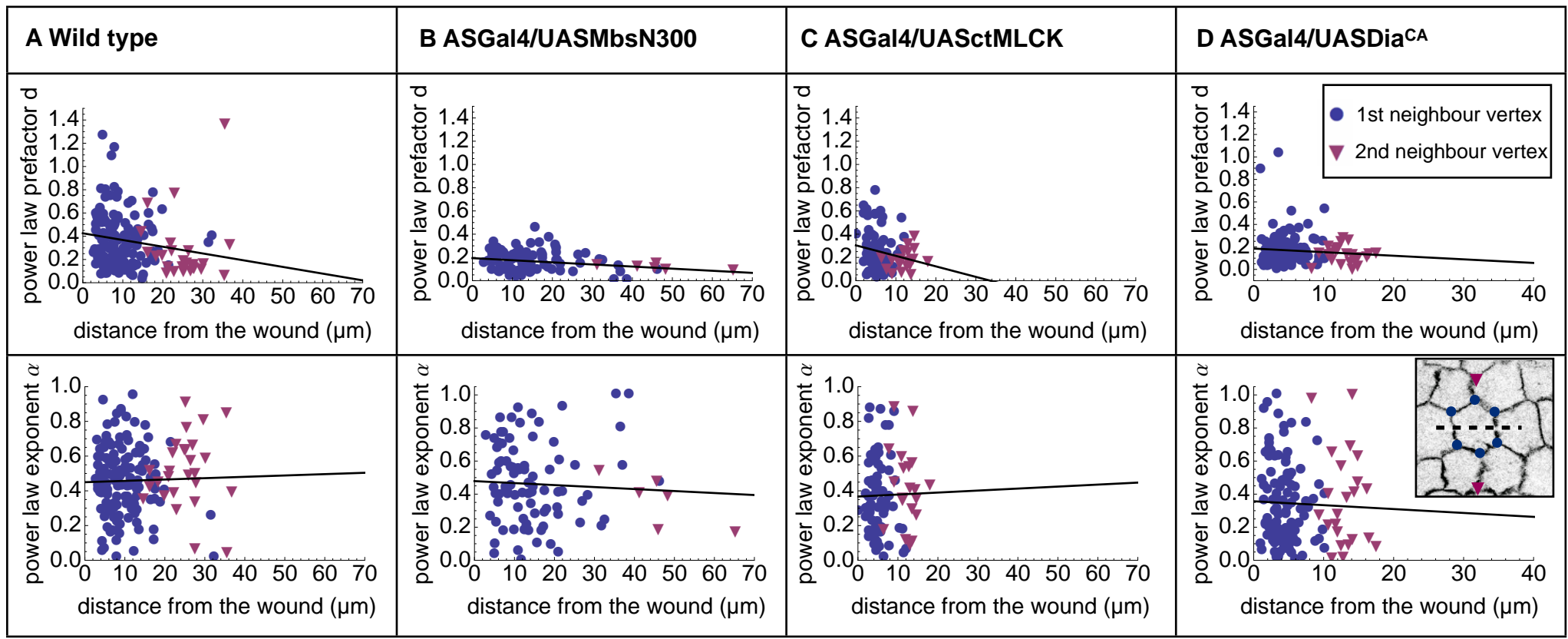
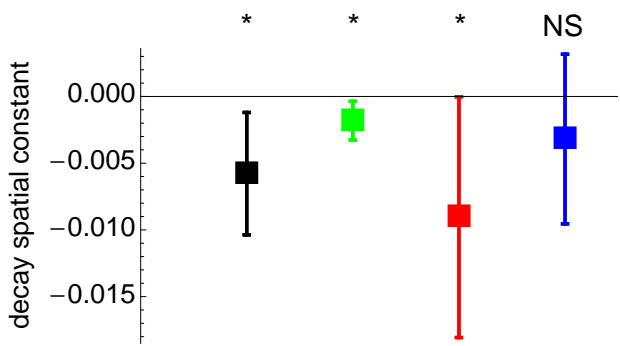
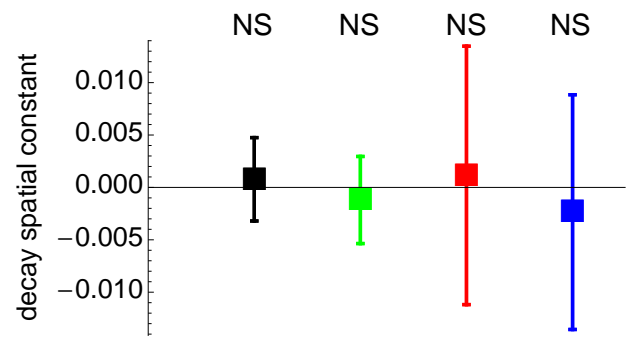

**Figure 4**

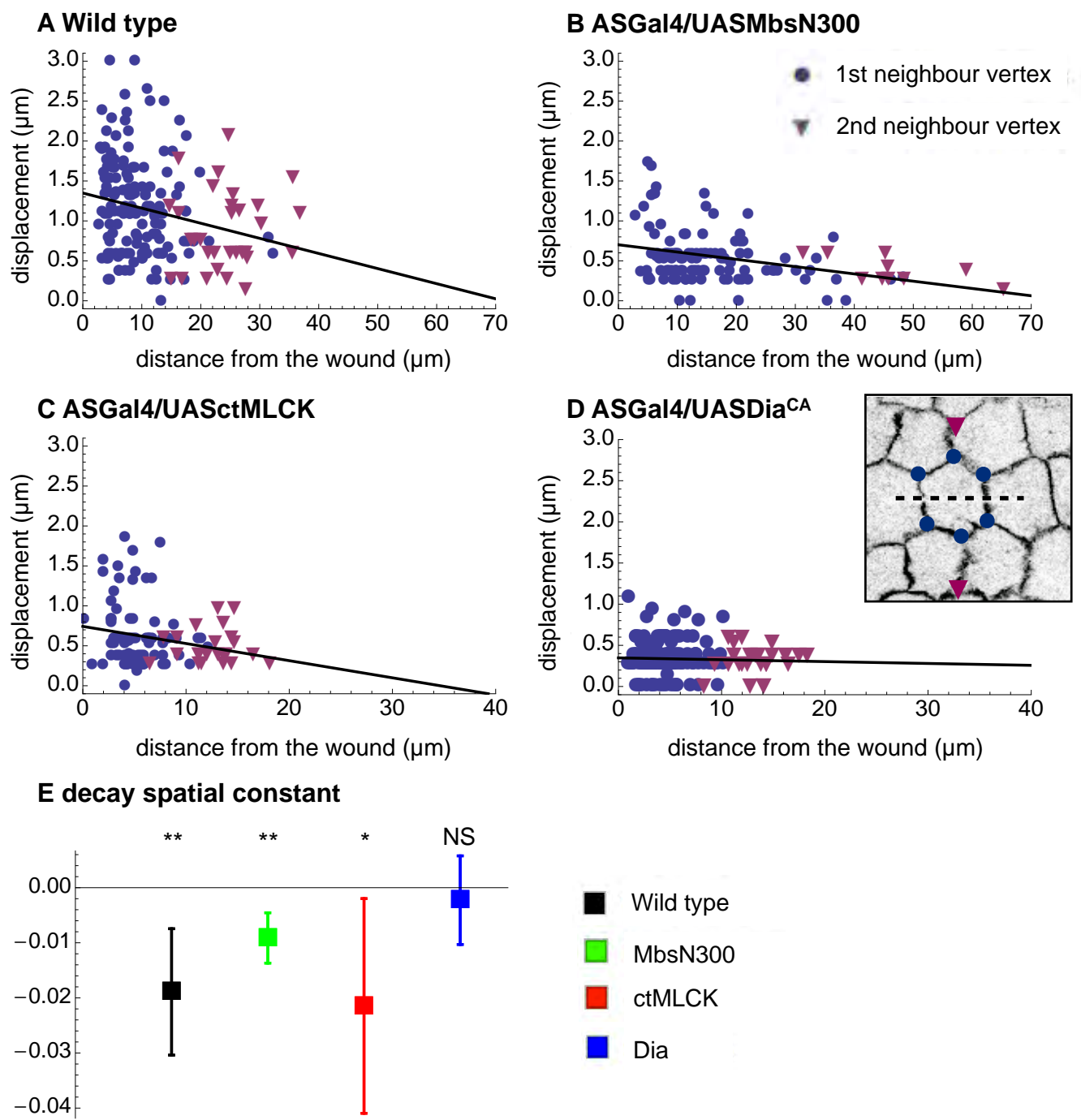

**Figure 5**

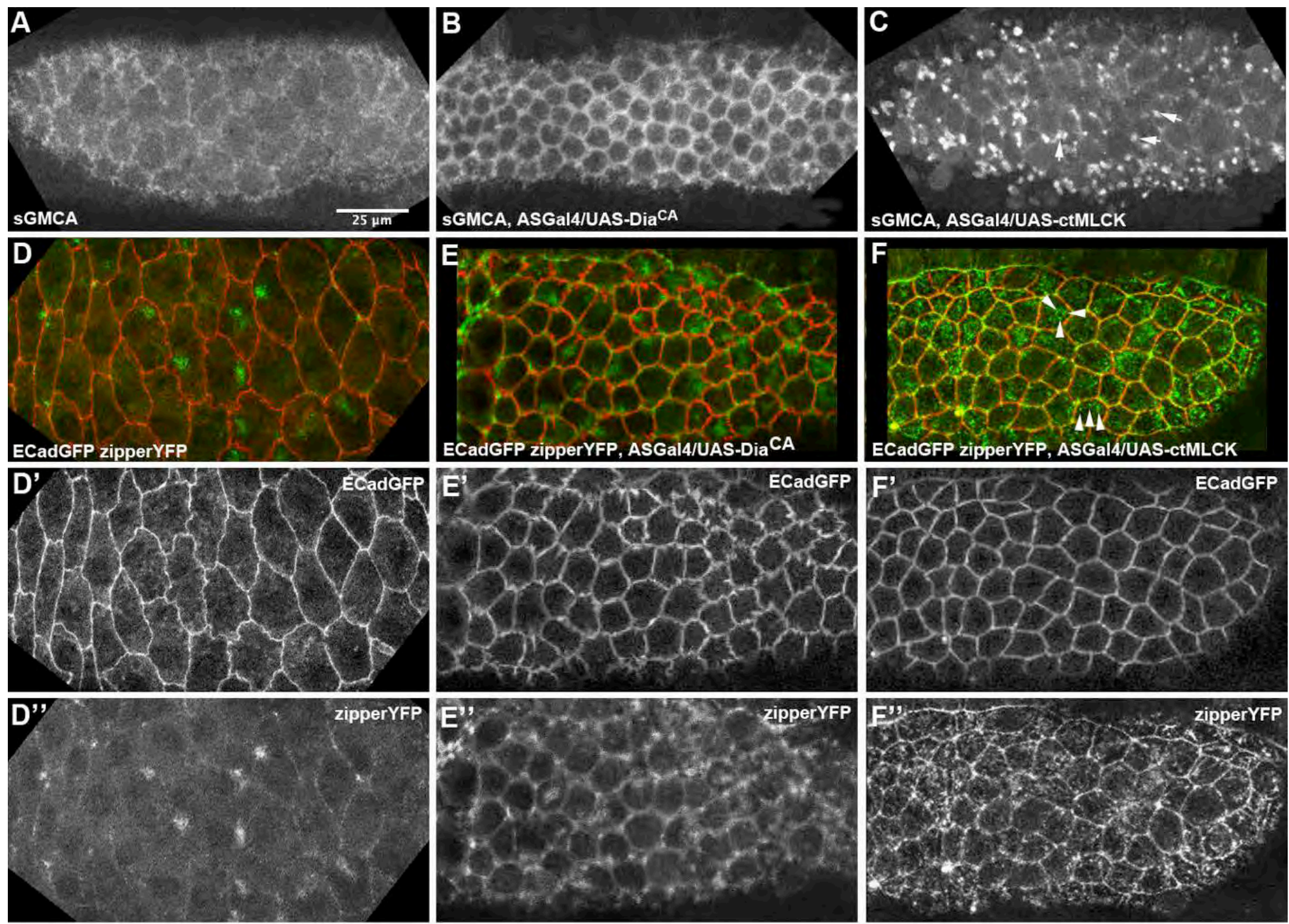

**Figure 6**

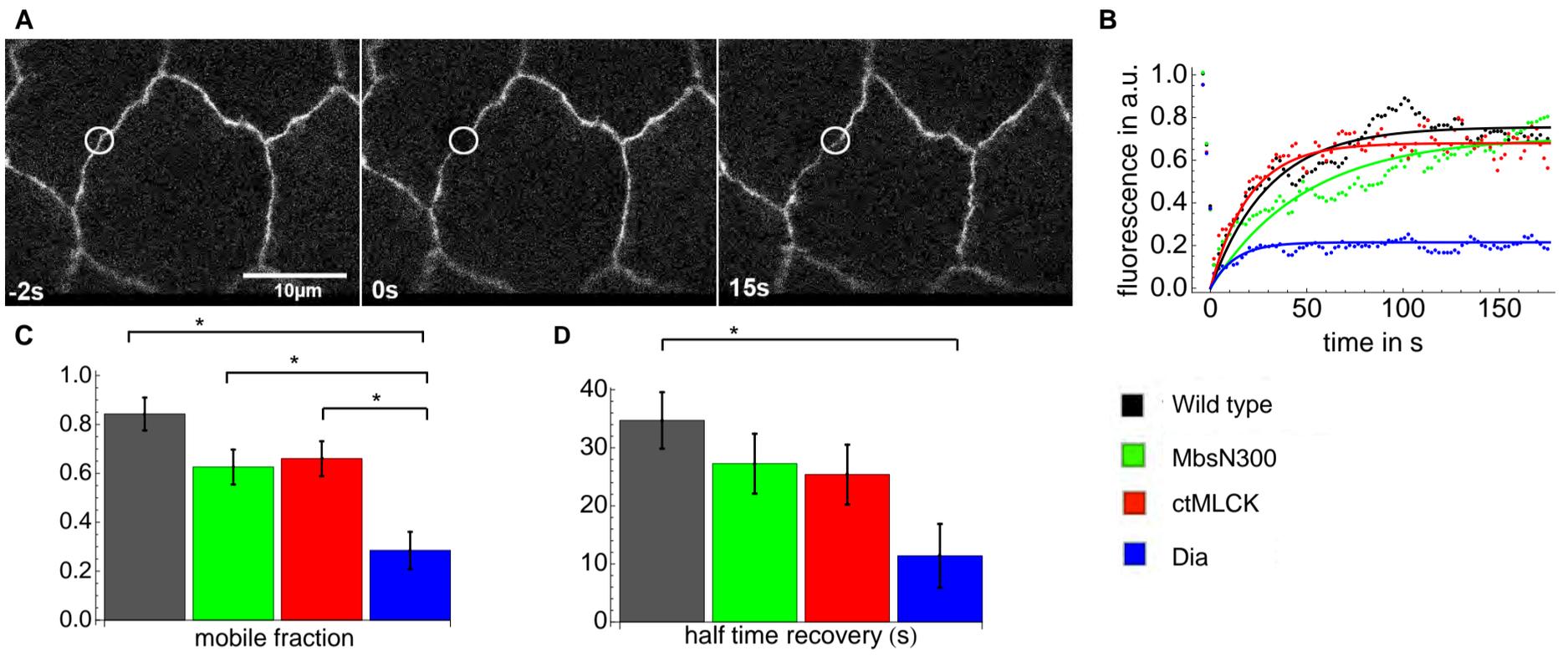

**Figure 7**

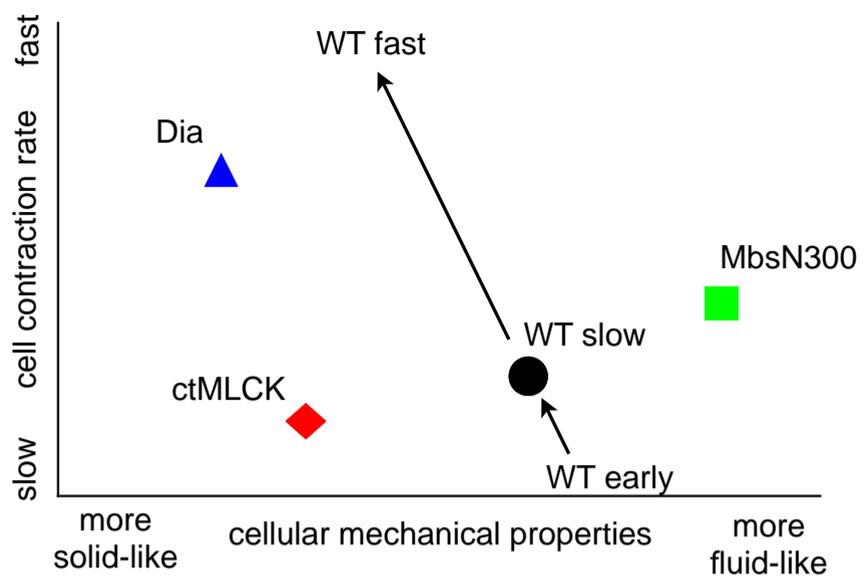

**Figure S1**

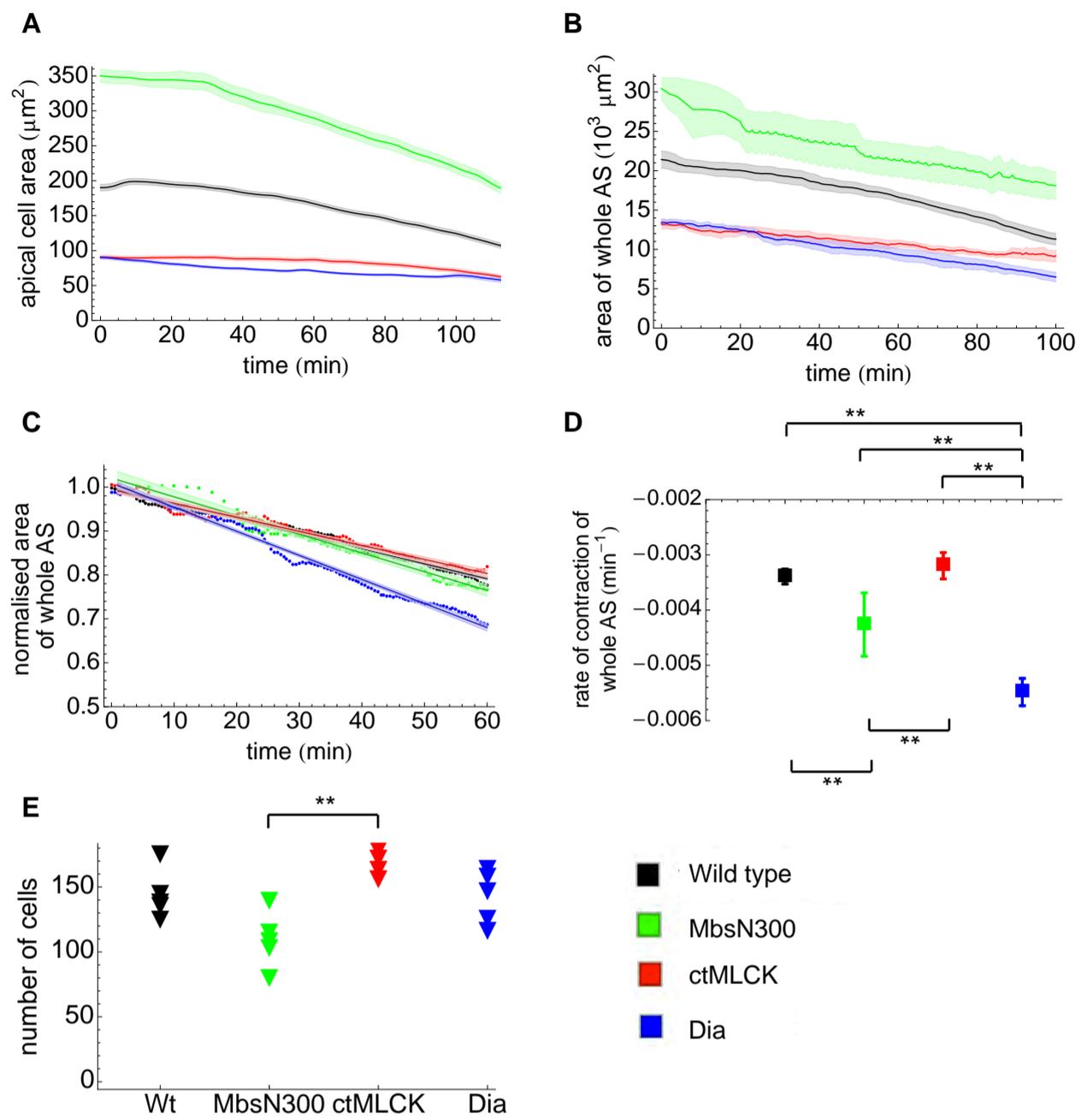

# Figure S2

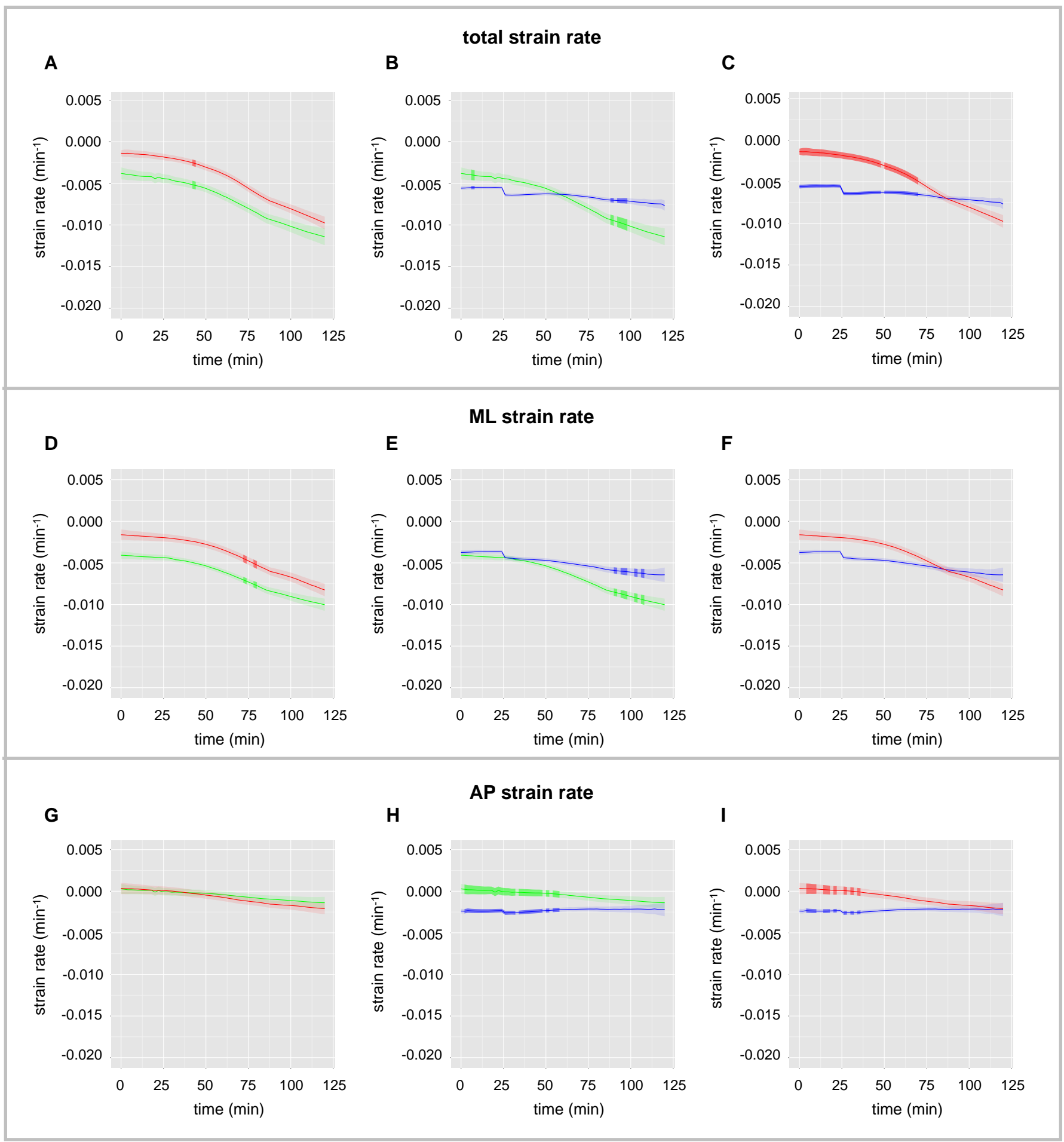

**Figure S3**

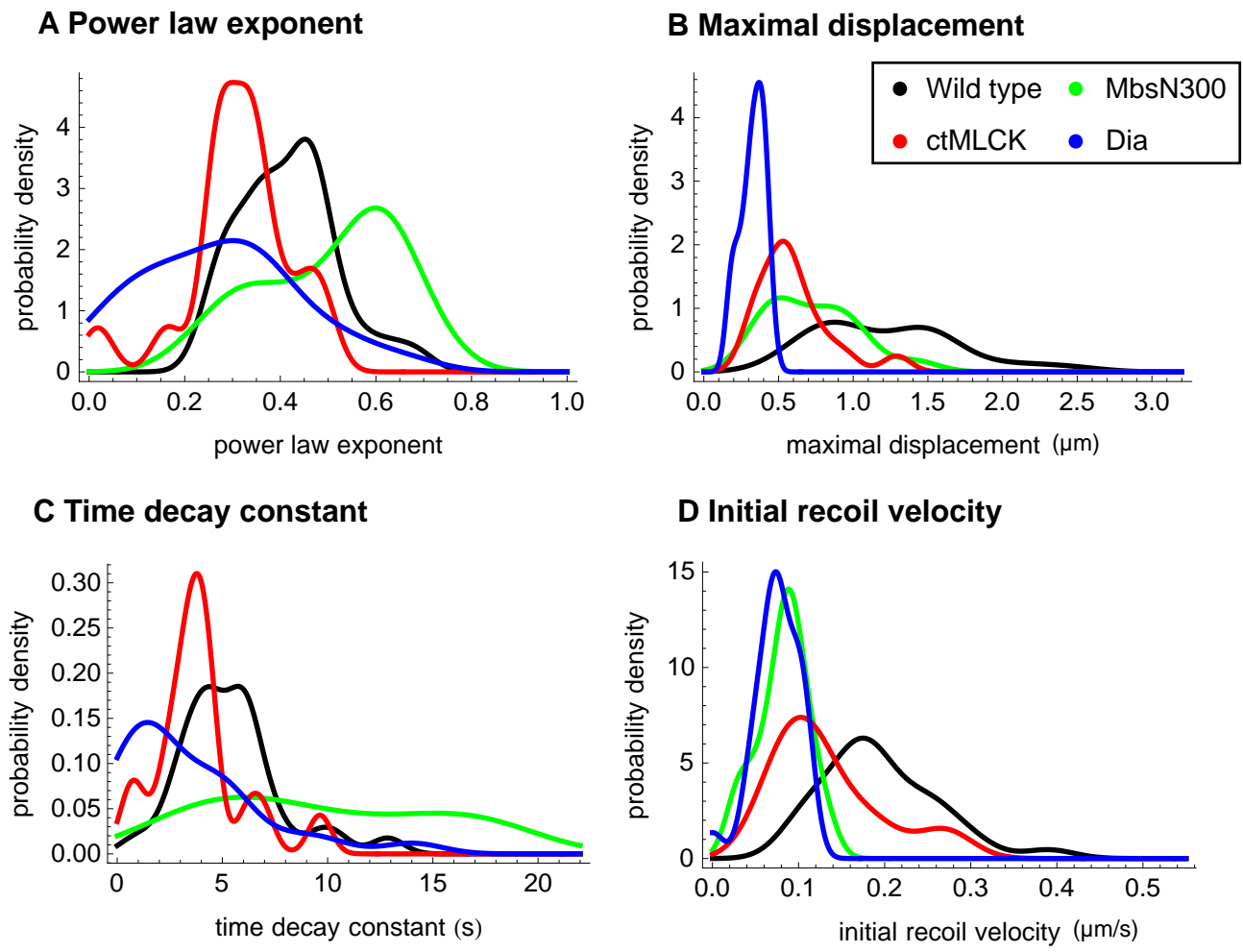

**Figure S4**

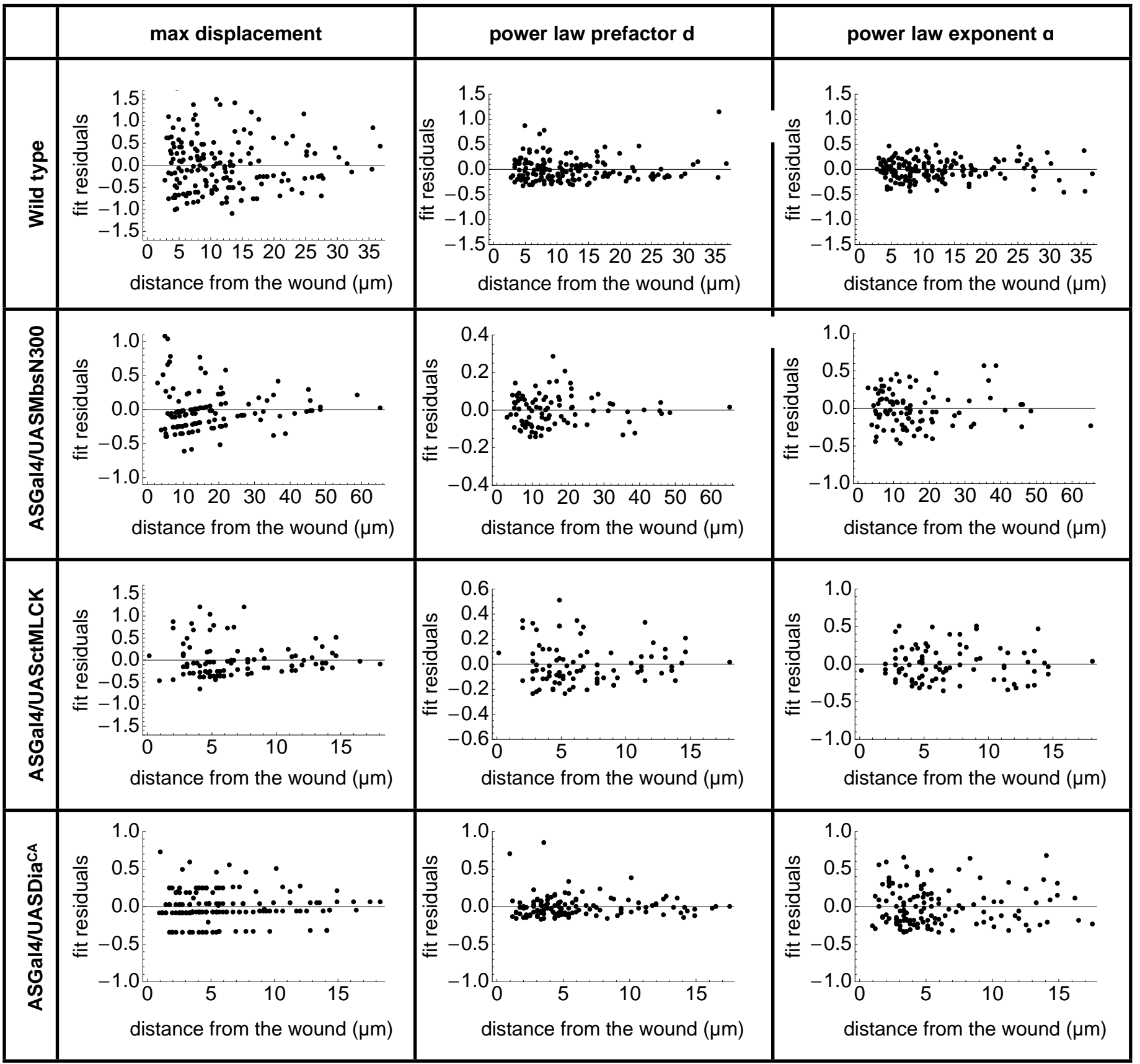

**Figure S5**

**A** ASGal4/UASDia<sup>CA</sup>, cut along AP

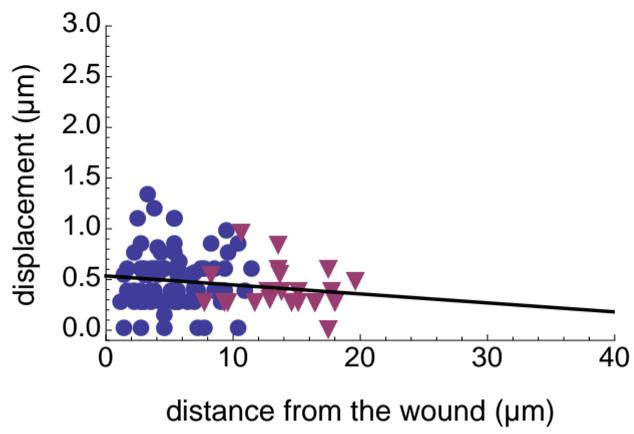

**B** ASGal4/UASDia<sup>CA</sup>, cut along ML

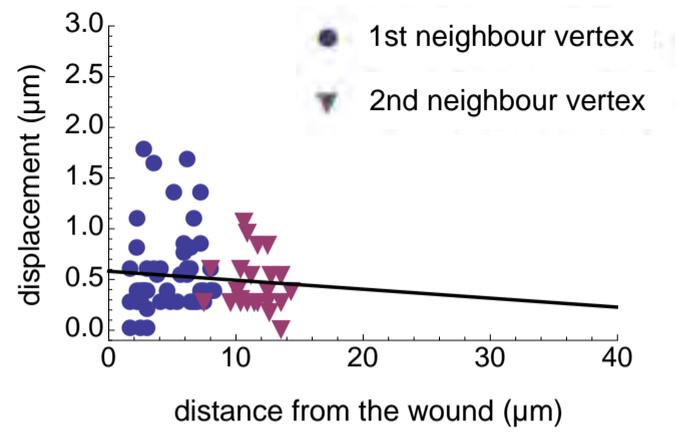

- 1st neighbour vertex
- 2nd neighbour vertex

**C** ASGal4/UASctMLCK, cut along AP

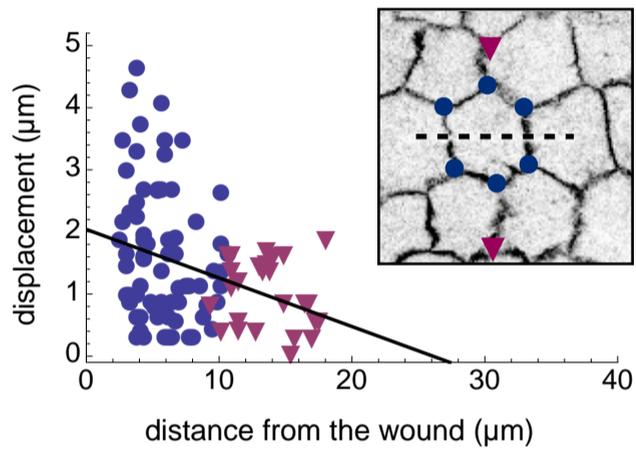

**D** decay spatial constant

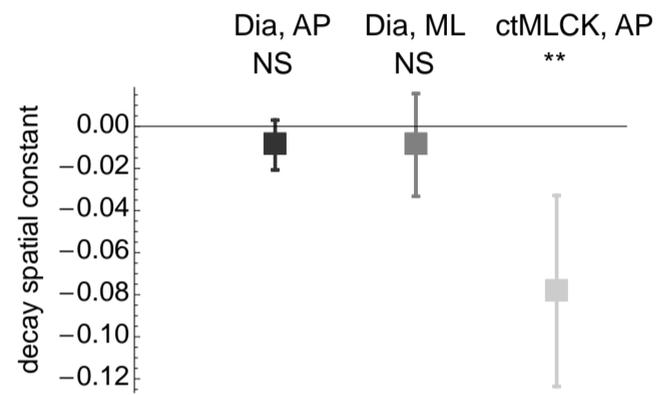

| Dia, AP | Dia, ML | ctMLCK, AP |
| NS | NS | ** |